\documentstyle[12pt]{article}
\newcommand{\be}{\begin{equation}}
\newcommand{\bea}{\begin{eqnarray}}
\newcommand{\eea}{\end{eqnarray}}
\newcommand{\ba}{\begin{array}}
\newcommand{\ea}{\end{array}}
\newcommand{\ee}{\end{equation}}

\expandafter\ifx\csname mathbbm\endcsname\relax

\else

\fi
\def\l{\label}
\def\o{\over}

\begin{document}
\begin{titlepage}
\hfill
\vbox{
    \halign{#\hfil         \cr
           hep-th/9703186 \cr
           IPM-97-198  \cr
           } 
      }  
\vspace*{3mm}
\begin{center}
{\LARGE Simple Derivation of the Picard-Fuchs equations for 
the Seiberg-Witten Models \\}
\vspace*{20mm}
{\ Mohsen Alishahiha }\\
\vspace*{1mm}
{\it Institute for Studies in Theoretical Physics and Mathematics, \\
 P.O.Box 19395-5531, Tehran, Iran } \\
{\it Department of Physics, Sharif University of Technology, \\
\it  P.O.Box 11365-9161, Tehran, Iran }\\
\vspace*{25mm}
\end{center}
\begin{abstract}
A closed form of the Picard-Fuchs equations for $N=2$
supersymmetric Yang-Mills theories with massless hypermultiplet are 
obtained for classical Lie gauge groups. We consider any number of massless
matter in fundamental representation so as to keep the theory
asymptotically
free.
\end{abstract}
\end{titlepage}
\newpage
\section{Introduction}

Recently duality has become a very important tool both in supersymmetric
Yang-Mills theories and string theory. Seiberg and Witten\cite{SW} 
have used duality and holomorphy to obtain the exact prepotential of $N=2$
SYM theory with gauge group $SU(2)$. (for review see: e.g.\cite{LER},
\cite{BIL} and \cite{HASS}). 

The key point in $N=2$ SYM models was the discovery of a hyperelliptic curve
with $r$ complex dimensional moduli space ($r$ is the rank of the 
guage group) with
certain singulareties, which gives information about the low energy
Willsonian effective action. The Seiberg-Witten data is a hyperelliptic
curve with a certain meromorphic one form ($E_{u_i},\lambda_{SW}$).

Indeed, the prepotential of $N=2$ SYM theory in the coulomb phase can be 
described with the aid of a family of complex curves with the identification 
of the vacuum expectation value (v.e.v) $a_i$ and its dual $a^D_i$
with the
periods of the curve
\be \l{AAD}
a_{i} = \oint _{\alpha _{i}} \lambda_{SW} \;\;\;\; \mbox{and} \;\;\;\;
a^{D}_{i} = \oint _{\beta_{i}} \lambda_{SW} , \label{integrals}
\ee
where $\alpha_{i}$ and $\beta_{i}$ are the homology cycles of the
corresponding Riemann surface.

There are two well known methods for finding the periods
and thereby the prepotential. The first method is to calculate the periods 
directly from the above integrals. This method has been developed in\cite{DHO}
and \cite{SAS}. They explicitly calculated the full expansion of the
renormalized order parameters using the method of residues. By this
method, they worked out explicitly the perturbative corrections as well
as the one and two instanton contributions to the effective prepotential.

On the other hand, one may use the fact that the periods $\Pi=(a_i,a^{D}_i)$
satisfy the Picard-Fuchs equations. Probably from the Picard-Fuchs equations
one can obtain the prepotential in an analytic way, which is for example,
important in the instanton calculus. Recently some
of these equations have been obtained in \cite{KL} 
and\cite{MU}. Also in \cite{ALI} we obtained a simple 
closed form of the Picard-Fuchs equations for Pure $N=2$ SYM theories 
for classical Lie gauge groups. The Picard-Fuchs equations, can also 
be obtained from the mirror symmetry in Calabi-Yau manifold\cite{VAFA}.

In this article we extend the results of \cite{ALI} to obtain a closed 
form for $N=2$ SYM theories with classical Lie gauge groups which have 
massless matter in fundamental representation.

The hyperelliptic curves for classical
gauge groups with any number of matter in the fundamental representation
are known \cite{PHO},\cite{HAN} and \cite{SHP}. Although in some cases
different hyperellieptic curves have been proposed for the same gauge 
group and the same hypermultiplet contents, but it was shown in \cite{DHP} 
by explicit calculations up to two instanton processes, that the 
corresponding effective prepotentials are the same for all 
these different curves. This equivalence results from the fact that the 
effective prepotential is unchanged under analytic reparametrizations 
of the classical order parameters\cite{DHP}.

The Seiberg-Witten data ($E_{u_i},\lambda_{SW}$) for classical gauge groups
with $n_f$ massless matter in fundamental representation have been proposed
as follows (see: \cite{PHO},\cite{HAN} and \cite{SHP})
\bea \l{SWD}
y^2&=&p^2(x)-G(x),\cr
& & \cr
\lambda_{SW}&=&({G'\o {2G}}p-p'){{x dx} \o y},
\eea
where
\be
p=x^{m+\epsilon}-\sum_{i=2}^m{u_i x^{m+\epsilon-i}}
\ee
with $m=r+1, \, i=2,3,...,\, \epsilon=0$ for $A_r$ series and
$m=2r, \, i=2,4,...,\, \epsilon=0$ for $B_r$ and $D_r$ series,
and $m=2r, \, i=2,4,...,\, \epsilon=2$ for $C_r$ series, and
$u_i$'s, the Casimirs of the gauge groups. Also
\be\ba {ll}
G=\Lambda^{2m-n_f} (x-\delta_{2m-1,n_f})^{n_f}\,\,\, & for\,\, A_r \cr
&\cr
G=\Lambda^{2m-2-2n_f} x^{2+2n_f}\,\,\, & for\,\, B_r \cr
&\cr
G=\Lambda^{2m+4-2n_f} x^{2n_f}\,\,\, & for\,\, C_r \cr
&\cr
G=\Lambda^{2m-4-2n_f} x^{4+2n_f}\,\,\, & for\,\, D_r. 
\ea\ee

Note that the $D_r$ series has an exceptional Casimir, $t$, of degree $r$,
but in our notation we set $u_{2r}=t^2$.
                                        
From explicit form of $\lambda_{SW}$ and the fact that the $\lambda_{SW}$
is lineary dependent on the Casimirs, setting ${\partial \o
{\partial{u_i}}}=
\partial_i$ one can see
\bea \l{PARTIAL}
\partial_i\lambda_{SW}&=&-{x^{m+\epsilon-i} \o y}dx+d(*), \cr
& & \cr
\partial_i\partial_j\lambda_{SW}&=&-{x^{2m+2\epsilon-i-j} \o y^3}p(x)dx+d(*).
\eea

The procedure of derivation of the Picard-Fuchs equations is to find proper
linear combinations of ${x^{m-i} \o y}dx$ and ${x^{2m-i-j} \o y^3}p(x)dx$
which give total derivative, then by integration from two sides and using
(\ref{AAD}) and (\ref{PARTIAL}), one can find second order differential
equations for the periods.

For example, from the second equation of (\ref{PARTIAL}) one can find 
the following identity for the periods ${\cal L}_{i,j;p,q}\Pi=0$ where
\be\ba {ll} \l{ID}
{\cal L}_{i,j;p,q}=\partial_i\partial_j-\partial_p\partial_q, & i+j=p+q
\ea\ee
\section{$B_r$ and $D_r$ Cases}

From (\ref{SWD}) the proposed hyperelliptic curve for these gauge 
groups with $n_f\leq m-k-1$
massless matter in the fundamental representation are
\be
y^2=p^2-\Lambda^{2m-2k-2n_f} x^{2k+2n_f},
\ee
where $k=1$ for $B_r$ and $k=2$ for $D_r$.

By direct calculation one can see that
\be \l{PLL2}
{d \o dx}({x^n \o y})=(n-k-n_f){x^{n-1} \o y}-(m-n_f-k){x^{m+n-1} \o y^3}p+
\sum_{i=2}^{m}{(m-k-n_f-i)u_i{x^{m+n-1-i} \o y^3}p}
\ee

Now from the equation (\ref{PARTIAL}), we can find the second order
diffrential equations for the periods (${\cal L}_n\Pi=0$) as follows
\be  \l{LL2}
{\cal L}_{n}=(k+n_f-n)\partial_{m-n+1}+(m-n_f-k)\partial_2\partial_{m-n-1}-
\sum_{i=2}^{m}{(m-k-n_f-i)u_i \partial_i\partial_{m-n+1}}.
\ee
Here $n=2s-1$ and $s=1,...,r-1$. Note that for $s=r$ equation (\ref{PLL2})
does not give the second order differential equation with respect to 
$u_i$. So by this method we can only obtain $r-1$ differential equations. 
An other equation can be obtained by following linear combination
\be
D_0=(k+n_f-m)d({x^{m+1} \o y})+\sum_{i=2}^{m}{(m-k-n_f+i)u_i
d({x^{m+1-i} \o y})}
\ee
or
\bea 
D_0&=&\lambda_{SW}-(\sum_{i=2}^{m}{i(i-2)u_i {x^{m-i} \o y}}
+\sum_{j,i=2}^{m}{ij u_iu_j {x^{2m-i-j} \o y^3}p}\cr 
&-&(m-n_f-k)^2 \Lambda^{2m-2n_f-2k} {x^{2n_f+2k} \o y^3}p)dx.
\eea

Note that, for the case of $n_f=m-k-1$, one can not write the last term in the
above equation as the form of ${x^{2m-i-j} \o y^3}p(x)dx$, so it only gives
the second order differention equation for the periods in the case of
$1 \leq n_f \leq m-k-2$, which is
\be \l{L2}
{\cal L}_r=1+\sum_{i=2}^{m}{i(i-2)u_i\partial_i}+\sum_{j,i=2}^{m}{ij u_iu_j
\partial_i\partial_j}-(m-n_f-k)^2 \Lambda^{2m-2k-2n_f}\partial^{2}.
\ee
where for $SO(m+1)$
\be\ba {ll}
\partial^2=\partial_{m-n_f-k}\partial_{m-n_f-k}\,\,\,\,& n_f\,\, odd\\
&\\
\partial^2=\partial_{m-n_f-2k}\partial_{m-n_f}\,\,\,\,& n_f\,\, even
\ea\ee
and for $SO(m)$
\be\ba {ll}
\partial^2=\partial_{m-n_f-k}\partial_{m-n_f-k}\,\,\,\,& n_f\,\, even\\
&\\
\partial^2=\partial_{m-n_f-k-1}\partial_{m-n_f-k+1}\,\,\,\,& n_f\,\, odd
\ea\ee

For the case of $n_f=m-k-1$ we should add an extra term to $D_0$ 
to cancel the last term, which is
\be
D=D_0+\Lambda^2 d({x^{m-1}\o y}).
\ee
so the last term of the (\ref{L2}) changes to
\be
\Lambda^2\sum_{i=2}^m {(i-1) u_i \partial_2\partial_i}.
\ee

Equations (\ref{LL2}) and (\ref{L2}) give a complet set of the 
Picard-Fuchs equations for the periods for gauge groups $B_r$ and $D_r$ with
any number of massless matter so as to keep the theory asymptoticlly free.
($n_f \leq m-k-1$).
\section{$C_r$ Case}
 
First let us write the Picard-Fuchs equatins for the pure gauge 
theory.\footnote{Here our notation for $C_r$ gauge group is 
difference from the one in\cite{ALI}}The proposed 
curve for pure $N=2$ SYM with gauge group $SP(m)$ is \cite{SHP}
\be
x^2y^2=p^2-\Lambda^{2m+4}
\ee
where 
\be
p=x^{m+2}-\sum_{i=2}^m{u_i x^{m+2-i}} + \Lambda^{m+2}\,\,\,i=2,4,...,m
\ee

By direct calculation one can see that
\be 
{d \o dx}({x^n \o z})=n{x^{n-1} \o z}-{{x^n p'}\o z^3}p,
\ee
where $z=xy$. So from (\ref{PARTIAL}) we have following second order
differential for the periods
\be  
{\cal L}_{n}=-n\partial_{m-n+3}+(m+2)\partial_2\partial_{m-n+1}-
\sum_{i=2}^{m}{(m+2-i)u_i \partial_i\partial_{m-n+3}}.
\ee
here $n=2s+1$ and $s=1,...,r-1$. As same as $B_r$ and $D_r$ cases, there
is an difficulty for $s=r$, again, we have only $r-1$ differential
equatins. One can see that, the last equation can be obtained 
by the following linear 
combination
\be
D_0=-(m+2)d({x^{m+3} \o z})+\sum_{i=2}^{m}{(m+2+i)u_i
d({x^{m+3-i} \o z})}-(m+2)^2\Lambda^{m+2}d({x\o z})
\ee
which gives a second order diffrential equation for the periods
\be 
{\cal L}_r=1+\sum_{i=2}^{m}{i(i-2)u_i\partial_i}+\sum_{j,i=2}^{m}{ij u_iu_j
\partial_i\partial_j}+(m+2)^2 \Lambda^{m+2}\sum_{i=0}^m{(m-i)u_i
\partial_m\partial_{i+2}}.
\ee
here $u_0=-1$.

Let us return to obtain the Picard-Fuchs equations for $C_r$ gauge
group with $n_f\leq m+1$ massless matter in the fundamental
representation. 
From (\ref{SWD}) the proposed curve for this theory is 
\be
y^2=p^2-\Lambda^{2(m+2-n_f)} x^{2n_f}
\ee
One can see that 
\be 
{d \o dx}({x^n \o y})=(n-n_f){x^{n-1} \o y}-(m+2-n_f){x^{m+n-1} \o y^3}p+
\sum_{i=2}^{m}{(m+2-n_f-i)u_i{x^{m+n-1-i} \o y^3}p}
\ee
wich gives the following differential equations
\be  
{\cal L}_{n}=(n_f-n)\partial_{m-n+3}+(m+2-n_f)\partial_2\partial_{m-n+1}-
\sum_{i=2}^{m}{(m+2-n_f-i)u_i \partial_i\partial_{m-n+3}}.
\ee

As in the previous cases from this method we obtain only $r-1$ differential
equations. The last equation can be obtained from the following linear 
combination
\be
D_0=(n_f-m-2)d({x^{m+3} \o y})+\sum_{i=2}^{m}{(m+2-n_f+i)u_i
d({x^{m+3-i} \o y})}
\ee
or
\bea 
D_0&=&\lambda_{SW}-(\sum_{i=2}^{m}{i(i-2)u_i {x^{m+2-i} \o y}}
+\sum_{j,i=2}^{m}{ij u_iu_j {x^{2m+4-i-j} \o y^3}p}\cr 
&-&(m+2-n_f)^2 \Lambda^{2m+4-2n_f} {x^{2n_f} \o y^3}p)dx.
\eea

Similar to $B_r$ and $D_r$ cases, for $n_f=1$ and $n_f=m+1$, the last term 
in the above expression can not rewrite in the form of
${x^{2m+4-i-j} \o y^3}p(x)dx$, so we should add an extra term to $D_0$.
For the case of $2\leq n_f \leq m$, the above equation gives a second 
order differential equation
\be \l{L3}
{\cal L}_r=1+\sum_{i=2}^{m}{i(i-2)u_i\partial_i}+\sum_{j,i=2}^{m}{ij u_iu_j
\partial_i\partial_j}-(m+2-n_f)^2 \Lambda^{2m+4-2n_f}\partial^2.
\ee
where $\partial^2=\partial_{m+2-n_f}\partial_{m+2-n_f}$ for even $n_f$ and
$\partial^2=\partial_{m+1-n_f}\partial_{m+3-n_f}$ for odd $n_f$. 
For $n_f=m+1$ $D_0$ should changes to
\be
D=D_0+\Lambda^2 d({x^{m+1}\o y}),
\ee
so the last term of the equation (\ref{L3}) changes to $\Lambda^2\sum{(i-1)
u_i\partial_2\partial_i}$, and for $n_f=1$
\be
D=D_0-{{(m+1)^2}\o u_m}\Lambda^{2(m+1)} d({x\o y}),
\ee
and the last term of the equation (\ref{L3}) changes to
\be
(m+1)^2 {\Lambda^{2m+2} \o u_m} \sum_{i=0}^{m-2}{(m+1-i)u_i\partial_{i+2}
\partial_m}
\ee
where $u_0=-1$.

\section{ $A_r$ Case}

Consider gauge group $SU(m)$ with $n_f\leq 2m-2$ massless 
hypermultiplets\footnote{For $n_f=2m-1$, because of $\Lambda$ dependent term
($(x-a_0\Lambda)^n_f$ where the coefficient $a_0$ comes from instanton
calculations), there is a difficulty, which also arises in the massive
case.
So we postpont it to future study.}
in the defining representation of the gauge group. From (\ref{SWD}) the 
hyperelliptic curve for this model is
\bea  \l{CU}
y^2&=&p^2(x)-\Lambda ^{2m-n_f}x^{n_f}
\eea
where
\bea
p(x)=x^m-\sum_{i=2}^{m}{u_i x^{m-i}}
\eea

By direct calculation one can see that
\be \l{PFA1}
{d \o dx}({x^n \o y})=(n-{n_f\o 2}){x^{n-1} \o y}-(m-{n_f\o 2})
{x^{m+n-1} \o y^3}p+
\sum_{i=2}^{m}{(m-{n_f\o 2}-i)u_i{x^{m+n-1-i} \o y^3}p}
\ee

From (\ref{PARTIAL}) we can find the second order
differential equation for the periods (${\cal L}_n\Pi=0$)
as follow
\be  \l{PF}
{\cal L}_{n}=({n_f\o 2}-n)\partial_{m-n+1}+(m-{n_f\o 2})\partial_2
\partial_{m-n-1}-
\sum_{i=2}^{m}{(m-{n_f\o 2}-i)u_i \partial_i\partial_{m-n+1}}.
\ee
where $n=s-1$ and $s=2,...,r-1$. As same as before, for $s=r$ (\ref{PFA1}) 
does not give the second order differential equation. Moreover, here
for $s=1$ same difficulty arises as well. So by this method we can only 
find $r-2$ equations. Two other equations can be obtaind by considering 
a particular linear combination of $d({x^j\o y})$.

Consider the following linear combination
\be
D_0=({n_f\o 2}-m)d({x^{m+1} \o y})+\sum_{i=2}^{m}{(m-{n_f\o 2}+i)u_i
d({x^{m+1-i} \o y})}
\ee
or
\be
D_0=\lambda_{SW}-(\sum_{i=2}^{m}{i(i-2)u_i {x^{m-i} \o y}}
+\sum_{j,i=2}^{m}{ij u_iu_j {x^{2m-i-j} \o y^3}p} -(m-{n_f\o 2})^2
\Lambda^{2m-n_f} {x^{n_f} \o y^3}p)dx.
\ee
which gives the second order differential equation for the periods in 
the case of $ 0\leq n_f \leq 2m-4$ and take the following form
\be  \l{PFA2}
{\cal L}_r=1+\sum_{i=2}^{m}{i(i-2)u_i\partial_i}+\sum_{j,i=2}^{m}{ij u_iu_j
\partial_i\partial_j}-(m-{n_f\o 2})^2 \Lambda^{2m-n_f}\partial^{2}.
\ee
where $\partial^{2}=\partial_{m}\partial_{m-n_f}$ for $n_f\leq m-2$ and
$\partial^{2}=\partial_{2}\partial_{m-l-2}$ and $l=-1,0,...,m-4$
for $ m-1 \leq n_f(=m+l)\leq 2m-4$.
For the cases of $ 2m-3 \leq n_f\leq 2m-2$, one should add an extra term
to $D_0$ as follow

For $n_f=2m-3$
\be \l{TTT1}
D=D_0+{3 \o 2} \Lambda^3 d({x^{m-2} \o y}),
\ee
so
\be \l{TTT2}
{\cal L}_r=1+\sum_{i=2}^{m}{i(i-2)u_i\partial_i}+\sum_{j,i=2}^{m}{ij u_iu_j
\partial_i\partial_j}+{3 \o 2}\Lambda^3\sum_{i=2}^m{(i-{3 \o 2})
u_i\partial_{3}\partial_i} +{3 \o 4}\Lambda^3\partial_3.
\ee

For $n_f=2m-2$
\be
D=D_0+\Lambda^2 d({x^{m-1} \o y}),
\ee
so
\be
{\cal L}_r=1+\sum_{i=2}^{m}{i(i-2)u_i\partial_i}+\sum_{j,i=2}^{m}{ij u_iu_j
\partial_i\partial_j}+\Lambda^2\sum_{i=2}^m{(i-1)
u_i\partial_{2}\partial_i}.
\ee

Finaly the last differential equation for the periods can be obtain from 
following linear combination (which is the analogous to $d({1 \o y})$ in
pure case \cite{ALI})
\be
E_0=(m-{n_f\o 2})d({x^m \o y})+{n_f \o 4}(n_f-2m-4)u_2 d({x^{m-2} \o y})
+(m-{n_f\o 2})^2\sum_{i=3}^{m}{u_i d({x^{m-i} \o y})}.
\ee
which gives second order differential equation for periods in the cases
of $1 \leq n_f\leq 2m-3$.
\bea  \l{PFA3}
{\cal L}_0 &=& c_2 (e-2)u_2\partial_3+\sum_{i=2}^m{(i-m)e^2 u_i\partial_{i+1}}
+\sum_{i=3}^m {c_{i}e u_i \partial_{i-1} \partial_{2}}
+\sum_{i=2}^m{c_2 (e-i)u_2u_i\partial_{3}\partial_i} \cr
&+& \sum_{i,j=2}^m{(i-m)e^2 u_iu_j \partial_{i+1}\partial_j}
+e^2{n_f \o 2} \Lambda^{2m-n_f}\partial^2.
\eea
where $\partial^2=\partial_m\partial_{m-n_f+1}$ for $n_f\leq m-1$ and
$\partial^2=\partial_2\partial_{m-l-1}$, $l=0,...,m-3$ 
for $m \leq n_f(=m+l)\leq 2m-3$ and also $e=(m-{n_f \o 2})$, 
$c_i=me+{in_f\o 2}$.
For the case of $n_f=2m-2$ one should add an extra term as follow
\be
E=E_0-(m-1)\Lambda^2 {d\o dx}({x^{m-2} \o y})
\ee 
and then the last term in equation (\ref{PFA3}) changes to
\be
(1-m)\Lambda^2(\partial_3 -\sum_{i=2}^m{(1-i)u_i\partial_3\partial_i}).
\ee

At the end, note that the equation (\ref{TTT2}) is not valid for $SU(2)$
with
$n_f=1$ as it should be. This incosistency comes from the $d({1 \o y})$ term in
(\ref{TTT1}). One can see that the Picard-Fuchs equation for this case
obtain
from the following combination
\be 
D=D_0+{9 \o 32}{\Lambda^3\o u}d({{x^2-3u}\o y}),
\ee
and gives well known results as follows
\be
{\cal L}_1=1+(4u^2+{27\o 64}{\Lambda^6 \o u})\partial^2_u.
\ee
 
\section{Conclusion}

To compare our results for the groups of rank $r\leq3$  with the 
results of \cite{MU} and \cite{KL}, let us for example consider
$SU(4)$ with one massless matter. From the equations (\ref{PFA1}), 
(\ref{PFA2}) and (\ref{PFA3}) we have
\bea
{\cal
L}_0&=&8u_2\partial_3+49u_3\partial_4-217u_3\partial_{22}+(8u_2^2-224
u_4)\partial_{23}+117u_2u_3\partial_{24}\cr
&+&(49u_3^2+128u_2u_3)\partial_{34}
+(49u_3u_4-{49 \o 2}\Lambda^7)\partial_{44},\cr
&& \cr
{\cal L}_1&=&\partial_4-7\partial_{22}+3u_2\partial_{24}+u_3\partial_{34}
-u_4\partial_{44},\cr
&&\cr
{\cal L}_3&=&1+3u_3\partial_3+8u_4\partial_4+4u_2^2\partial_{22}
+(9u_3^2+16u_2u_4)\partial_{33}+16u_4^2\partial_{44}+12u_2u_3\partial_{23}\cr
&+&(24u_3u_4-{49\o4}\Lambda^7)\partial_{34},
\eea
where $\partial_{ij}=\partial_i\partial_j$. One can check that these
equations
are linear combination of those of\cite{MU}.

To summarize, we have obtained a closed form of the Picard-Fuchs equatins
for $N=2$ SYM theories with classical Lie gauge groups which have
massless matter in fundamental representation.

Note: After completion of this work, I received paper\cite{SCH} which
has considerable overlap with our work.

{\bf Acknowledgement}

I would like to thank S. Randjbar-Daemi for uesful commends.

\end{document}